\newcounter{myctr}
\def\myitem{\refstepcounter{myctr}\bibfont\noindent\ifnum\themyctr>9\else\phantom{0}\fi\hangindent17pt\themyctr.\enskip}

\documentclass{ws-ijqi}

\begin{document}

\markboth{Laura Mazzola, Jyrki Piilo and Sabrina Maniscalco}
{Frozen discord in non-Markovian dephasing channels}

%%%%%%%%%%%%%%%%%%%%% Publisher's Area please ignore %%%%%%%%%%%%%%
\catchline{}{}{}{}{}
%%%%%%%%%%%%%%%%%%%%%%%%%%%%%%%%%%%%%%%%%%%%%%%%%%%%%%%%%%%%%%%%%%%

\title{FROZEN DISCORD IN NON-MARKOVIAN DEPHASING CHANNELS}

\author{LAURA MAZZOLA\footnote{laumaz@utu.fi}}

\address{Turku Centre for Quantum Physics, Department of Physics
and Astronomy, University of Turku, FI-20014 Turun yliopisto,\\
Turku, Finland.\\
\ laumaz@utu.fi}

\author{JYRKI PIILO}

\address{Turku Centre for Quantum Physics, Department of Physics
and Astronomy, University of Turku, FI-20014 Turun yliopisto,\\
Turku, Finland.\\
\ jyrki.piilo@utu.fi}

\author{SABRINA MANISCALCO}

\address{Turku Centre for Quantum Physics, Department of Physics
and Astronomy, University of Turku, FI-20014 Turun yliopisto,\\
Turku, Finland.\\
\ smanis@utu.fi}

\maketitle

\newcommand{\ket}[1]{\displaystyle{|#1\rangle}}
\newcommand{\bra}[1]{\displaystyle{\langle#1|}}
\newcommand{\al}{\alpha}
\newcommand{\Om}{\Omega}
\newcommand{\om}{\omega}
\newcommand{\D}{\Delta}
\newcommand{\G}{\Gamma}
\newcommand{\g}{\gamma}
\newcommand{\s}{\sigma}
\newcommand{\la}{\lambda}

\begin{history}
\received{Day Month Year}
\revised{Day Month Year}
%\accepted{Day Month Year}
%\comby{(xxxxxxxxxx)}
\end{history}

\begin{abstract}
We investigate the dynamics of quantum and classical correlations in
a system of two qubits under local colored-noise dephasing
channels. The time evolution of a single qubit interacting with its
own environment is described by a memory kernel non-Markovian master
equation. The memory effects of the non-Markovian reservoirs
introduce new features in the dynamics of quantum and classical
correlations compared to the white noise Markovian case. Depending
on the geometry of the initial state the system can exhibit frozen
discord and multiple sudden transitions between classical and
quantum decoherence [L. Mazzola, J. Piilo, S. Maniscalco,
Phys.~Rev.~Lett.~\textbf{104}, 200401 (2010)]. We provide a
geometric interpretation of those phenomena in terms of the distance
of the state under investigation to its closest classical state in
the Hilbert space of the system.
\end{abstract}


\section{Introduction}

In the last two decades lots of interest have been devoted to the
definition and understanding of correlations in quantum systems.
From the first formalization of the separability problem by Werner
\cite{Werner} a number of fundamental results have been found in
this field. In particular recently a big deal of attention has been
devoted to the definition and study of quantum and classical
correlations in quantum systems. The works of Ollivier and Zurek
\cite{Ollivier}, and Henderson and Vedral \cite{Henderson} had marked
the beginning of a new line of research shifting the attention from
the entanglement vs.~separability dichotomy to the quantum
vs.~classical paradigm. The fact that separable states can be prepared
via local operations and classical communications between two
parties has often been interpreted as a signature of classicality.
However, there are quantum correlations which are not captured by
entanglement. For example the state
$\rho=1/2(\ket{0}\bra{0}_{A}\otimes\ket{-}\bra{-}_{B}+\ket{+}\bra{+}_{A}\otimes\ket{1}\bra{1}_{B})$
with $\ket{\pm}=(\ket{0}\pm\ket{1})/\sqrt{2}$ is separable,
nevertheless it cannot be described by classical means, i.e., by a
classical bivariative probability distribution. The reason for that
is hidden in the non-orthogonality of the states of $A$ and $B$, and
consequently in the impossibility to locally distinguish the states
of each subsystem.

Many different measures have been proposed to describe such
more-general-than-entanglement quantum correlations
\cite{Henderson,Opp,Grois,LuoMID,Modi}, among them probably the most
popular one is the quantum discord \cite{Ollivier}. The discord,
defined as the difference between the quantum generalization of two
equivalent formulations of classical mutual information, involves an
optimization procedure over the set of measurements on a given
subsystem. Therefore mathematical investigations have been carried
to find analytical expressions for the discord in different types of
systems, as in qubits \cite{Luo,Ali} and harmonic oscillators
\cite{Paris,Adesso}. The set of zero discord states has been shown to
be measure zero and nowhere dense \cite{Ferraro}, moreover a new
class of multipartite separable states, the pseudo-classical states,
has been introduced and showed to have interesting properties
\cite{Vacanti}. Proposals to witness the presence of quantum discord
have appeared in Ref. \cite{Dakic,Bylicka}. The role of quantum
discord has been investigated in quantum information tasks as the
power-of-one-qubit protocol \cite{Datta} or in the Grover search
algorithm \cite{Grover}, posing the possibility of using quantum
discord as a new, more general than entanglement, quantum
computation resource. The dynamics of quantum discord has been
studied even in quantum biology, specifically in light-harvesting
complex \cite{quantumbio}.

A very active area of investigation of the discord is its behaviour
under decoherence, i.e. the dynamics of quantum and classical
correlations in open quantum systems
\cite{Zurek03,Horodec,Rodriguez,Piani,Maziero0,Datta2,Piani2,Werlang,Maziero,Fanchini,Vasile,Jin-Shi}.
In particular in Ref. \cite{Mazzola} we have discovered that quantum
correlations, quantified by quantum discord, can be completely
unaffected by decoherence for long intervals of time. Specifically,
we have found that an open system of two qubits, in Markovian
dephasing channels can undergo, what we have called a sudden
transition between classical and quantum decoherence, namely a
transition between a \lq\lq classical decoherence\rq\rq phase in
which quantum correlation is frozen, while classical correlation is
lost, and a \lq\lq quantum decoherence\rq\rq regime in which quantum
correlation is deteriorated while classical correlation remains
constant.

Here we study the time evolution of classical and quantum
correlations for a system of two identical qubits under local
non-dissipative non-Markovian channels. We use a well established
model describing the dynamics of a two-level system in a
dephasing channel in which white noise is replaced by colored
noise \cite{Daffer}. There, the interaction with the environment is
schematized by random telegraph signal noise. We study the
non-trivial effects the non-Markovianity of the system brings to
light. As in the Markovian case, there exists a class of initial
states for which the system exhibits the sudden transition between
classical and quantum decoherence and the frozen
discord \cite{Mazzola,Jin-Shi}. However, here multiple transitions
can arise due to the memory effects of the environment.

The transition between the two dynamical regimes can be explained,
in the light of the results of Modi \textit{et al.}~\cite{Modi}.
There, the discord is defined as the relative entropy with respect to the closest
classical state, therefore frozen discord corresponds to a dynamical
evolution in which the state under investigation maintains a fixed
distance (relative entropy or trace distance) to its closest
classical state. In this framework the sudden
transition corresponds to a sudden change in the location of the closest
classical state in the Hilbert space of the system.

The outline of the paper is the following. In Sec.~II we introduce the
physical system under investigation \cite{Daffer}. In Sec.~III we
recall the concepts of quantum discord and classical correlation and
study the dynamics of those quantities in our specific system. In
Sec.~IV we illustrate the interpretation of the phenomenon
of frozen discord in terms of the distance to
the closest classical state set. Section V summarizes and concludes the paper.

\section{The model}

We begin by introducing the physical model under study: two
non-interacting qubits under local identical colored noise
dephasing channels. Since the two qubits are locally interacting
with their own environments, and are not coupled between each other,
they have independent dynamical evolutions. Therefore we can first
consider the dynamics of the single qubit, and then derive the
dynamics of the composite two-qubit system.

A generic memory kernel master equation can be written in the form
\begin{equation}\label{ME}
\dot{\rho}=K\mathcal{L}\rho,
\end{equation}
where $\rho$ is the density operator of the small system of
interest, in our case a qubit, $\mathcal{L}$ is a Lindblad
superoperator describing the dynamics induced by the interaction
with the environment, $K$ is an integral operator acting in the
following way $K\phi=\int_{0}^{t}k(t-t')\phi(t')dt'$, and $k(t-t')$
is the kernel function determining the type of memory of the
reservoir.

A master equation of this form arises when considering a two-level
system subjected to random telegraphic noise. A model like this
describes, for example, a spin in presence of a magnetic field,
having constant intensity but inverting its sign randomly in time.
It is possible to write a time-dependent phenomenological
Hamiltonian for this kind of system:
\begin{equation}\label{Hamiltonian}
H(t)=\hbar\sum_{i=1}^{3}\G_{i}(t)\sigma_{i}
\end{equation}
here $\sigma_{i}$ are the Pauli operators and $\G_{i}(t)$ are
independent random variables. We consider the general case of noise
in the three directions. Each random variable can be written as
$\G_{i}(t)=a_{i}n_{i}(t)$. The random variable $n_{i}(t)$ has a
Poisson distribution with a mean equal to $t/2\tau_{i}$, while
$a_{i}$ is a coin-flip random variable assuming the values $\pm
a_{i}$. By considering the von Neumann equation
$\dot{\rho}=-(\imath/\hbar)[H,\rho]$, one can find a formal solution
for the qubit density matrix operator of the form
\begin{equation}\label{formalsol}
\rho(t)=\rho(0)-\imath\int_{0}^{t}\sum_{k}\G_{k}(s)[\sigma_{k},\rho(s)]ds.
\end{equation}
When inserting such a formal solution back into the von Neumann
equation and performing a stochastic average, one obtains the
following memory kernel master equation
\begin{equation}\label{memorykernelME}
\dot{\rho}(t)=-\int_{0}^{t}\sum_{k}\exp{(-(t-t')/\tau_{k})}a_{k}^2[\sigma_{k},[\sigma_{k},\rho(t')]dt'
\end{equation}
in which the memory kernel comes from the correlation functions of
the random telegraph signals
$\langle\G_{j}(t)\G_{k}(t')\rangle=a_{k}^2\exp{\left(-|t-t'|/\tau_{k}\right)}\, \delta_{jk}$.

The dynamics of this system and in particular the conditions of
complete positivity of the map corresponding to such a master
equation have been studied in detail by Daffer {\it et al.}~in
Ref.~\cite{Daffer}. In particular they demonstrated that complete
positivity is assured when two of the $a_{i}$ are zero, representing
the physical situation of noise in only one direction. This is the
case we are focusing here. In that work they also provided a Kraus
operator representation of the map
$\Phi_{t}(\rho)=\sum_{k}A_{k}^{\dag}\rho A_{k}$. Depending on which
of the $a_{i}=a$ is non zero the map has Kraus operators
\begin{equation}\begin{split}\label{Lambdas}
A_{i}=&\sqrt{[1-\Lambda(\nu)]/2}\sigma_{i},\\
A_{j}=&0,\quad A_{k}=0,\\
A_{4}=&\sqrt{[1+\Lambda(\nu)]/2}I,
\end{split}\end{equation}
where $\Lambda(\nu)=\exp{\nu}[\cos{(\mu\nu)}+\sin{(\mu\nu)}/\mu]$,
$\mu=\sqrt{(4a\tau)^2-1}$, $\nu=t/2\tau$ is a dimensionless time,
$i=1,\ 2,\ 3$ is the direction of noise, and $j$ and $k$ the directions
in which there is no noise. So by changing the direction of the
noise we get a colored noise bit flip, bit-phase flip or phase flip
channel, respectively. Having the Kraus operators of a single qubit,
one can write the evolution of the generic state of the two qubits
each interacting locally with its own environment as
\begin{equation}\label{Phi}
\Phi_{t}(\rho_{AB})=\sum_{i,j}A_{i}^{(A)}A_{j}^{(B)}\rho_{AB}A_{i}^{(A)\dag}A_{j}^{(B)\dag},
\end{equation}
where $\Phi_{t}(\cdot)$ is the completely positive trace preserving
map ruling the evolution of the system.

It is worth mentioning that, even though in general the mathematical structure of the
memory kernel master equations in Eq. \eqref{memorykernelME} do not
guarantee the presence of non-Markovian features in the dynamics
\cite{memkern}, however the quantum process described by
$\Phi_{t}(\cdot)$ is non-Markovian according to the measure of
non-Markovianity of Ref.~\cite{NMmeasure}.

\section{Dynamics of quantum and classical correlations}
The physical quantities we want to investigate here are the quantum
and classical correlations between the two qubits.
%As announced in the introduction,
We use the quantum discord as a measure of quantum
correlations.

The idea behind quantum discord is to exploit the difference between
quantum extensions of classically equivalent concepts to evaluate
the \lq\lq quantumness\rq\rq of a quantum system. In particular in
classical information theory there are two equivalent expressions
for the mutual information of a bipartite system. However, when
pursuing the quantum analogue, these two formulations differ, and one
can use the mismatch between the two quantum extensions of classical
mutual information to assess quantum correlations.

The first quantum generalization of mutual information is the so
called quantum mutual information:
\begin{equation}\label{mutual}
I(\rho_{AB})=S(\rho_{A})+S(\rho_{B})-S(\rho_{AB}),
\end{equation}
where $\rho_{AB}$ is the density matrix of the total system,
$\rho_{A(B)}$ is the reduced density matrix of subsystem $A(B)$, and
$S(\rho)=-\rm{Tr}\{\rho \log_{2}\rho\}$ is the von Neumann entropy. This
is generally accepted as the measure for the total amount of
correlations (quantum and classical) of a quantum system
\cite{Grois,Schumacher}.

The second extension of mutual information requires the
generalization of the concept of conditional entropy. Indeed,
performing measurements on system B affects our knowledge of system
A, in particular how much system A is modified by the measurement depends on the choice
of the measurement performed on B. Here the measurement ${B_{k}}$ is
considered of von Neumann type and it is described by a complete set
of orthonormal projectors ${\Pi_{k}}$ corresponding to outcome k. So
the conditional density operator, which is the quantum state of the
total system conditioned on the measurement outcome labeled by k,
becomes $\rho_{k}=(I\otimes B_{k})\rho_{AB} (I\otimes B_{k})/p_{k}$ where
$p_{k}=\rm{Tr}\{(I\otimes B_{k})\rho_{AB}(I\otimes B_{k})\}$, and $I$ is the
identity operator for subsystem A. Defining the quantum analog of
conditional entropy as
$S(\rho_{AB}|\{B_{k}\})=\sum_{k}p_{k}S(\rho_{k})$ we can introduce
the second quantum extension of mutual information.
\begin{equation}\label{ext2mutual}
J(\rho_{AB}|{B_{k}})=S(\rho_{A})-S(\rho_{AB}|{B_{k}}).
\end{equation}
In Ref.~\cite{Henderson} Henderson and Vedral have shown that the
maximum of this quantity over all the possible set of
measurements can be interpreted as a measure of classical
correlations of the state
\begin{equation}\label{classical}
C(\rho_{AB})\equiv \max_{\{\Pi_{j}\}}[J(\rho_{AB}|{B_{k}})].
\end{equation}
Therefore, the difference between the the quantum mutual information
$I(\rho_{AB})$, describing total correlations, and the
measurement-based definition of quantum mutual information
$C(\rho_{AB})$, measuring classical correlations, defines the
so-called quantum discord
\begin{equation}\label{QD}
D(\rho_{AB})=I(\rho_{AB})-C(\rho_{AB}).
\end{equation}

In the following we investigate the dynamics of these quantities in
our non-Markovian dynamical model for a particular class of states.
The states we consider have maximally mixed marginals
%(meaning that
%the density matrices of the reduced system are maximally mixed), i.e.,
and have the form
\begin{equation}\label{maxmixedmarg}
\rho_{AB}=\frac{1}{4}\left(1_{AB}+\sum_{i=1}^{3}c_{i}\sigma_{i}^{A}\sigma_{i}^{B} \right),
\end{equation}
where $\sigma_{i}^{A(B)}$ are the Pauli operators in direction $i$
acting on A (B),  $c_{i}$ are real numbers such that
$0\leq|c_{i}|\leq1$ for every $i$, and $1_{AB}$ the identity operator
of system AB. This class of states can equivalently be written in
the form of Bell diagonal states
\begin{eqnarray}\label{Belldiag}
\rho_{AB}(t)&=&\la_{\Psi}^+(t)\ket{\Psi^{+}}\bra{\Psi^{+}}+\la_{\Phi}^+(t)\ket{\Phi^{+}}\bra{\Phi^{+}} \nonumber \\
&+&\la_{\Phi}^-(t)\ket{\Phi^{-}}\bra{\Phi^{-}}+\la_{\Psi}^-(t)\ket{\Psi^{-}}\bra{\Psi^{-}},
\label{evolution}
\end{eqnarray}
where
\begin{eqnarray}
\la_{\Psi}^{\pm}(t)&=&[1 \pm c_{1}(t) \mp c_{2}(t)+c_{3}(t)]/4, \label{la0} \\
\la_{\Phi}^{\pm}(t)&=&[1\pm c_{1}(t) \pm
c_{2}(t)-c_{3}(t)]/4,\label{la3}
\end{eqnarray}
and $\ket{\Psi ^{\pm}}=(\ket{00}\pm\ket{11})/\sqrt{2}$,
$\ket{\Phi^{\pm}}=(\ket{01}\pm\ket{10})/\sqrt{2}$ are the four Bell
states.

The expressions of the classical and quantum correlations for this
class of states are given by \cite{Luo}
\begin{equation}
C(\rho_{AB})=\sum_{k=1}^{2}\frac{1+(-1)^{k}\chi}{2}\log_{2}(1+(-1)^{k}\chi)
\label{cl2}
\end{equation}
\begin{equation}
D(\rho_{AB})=2+\sum_{k,l}\lambda_{k}^l(t)\log_{2}\lambda_{k}^l(t)-C(\rho_{AB}),
\end{equation}
where $\chi(t)=\max\{|c_{1}(t)|,|c_{2}(t)|,|c_{3}(t)|\}$,
$k=\Psi,\Phi$, and $l=\pm$.

Since the dynamical evolution map $\Phi_{t}(\cdot)$ of Eq.
\eqref{Phi} does not change the form of the state in Eq.
\eqref{maxmixedmarg}, classical and quantum correlations are given
by those expressions throughout all the time evolution.

As already noticed at the end of Sec.~II, the interaction with
the environment gives rise to a bit flip, bit-phase flip or phase
flip channel when the direction of the noise is along x, y
or z (in Eq. \eqref{Lambdas} $i=1,2,3$), respectively. Using the
Kraus operators in Eq. \eqref{Lambdas} and the dynamical evolution
of the state in Eq. \eqref{Phi} we find that the parameters
$c_{i}(t)$ evolve in the following way
\begin{equation}
c_{i}(\nu)=c_{i}(0),\quad c_{j}(\nu)=c_{j}(0)\Lambda(\nu)^2,\quad
c_{k}(\nu)=c_{k}(0)\Lambda(\nu)^2,
\end{equation}
where $\nu=t/2\tau$ is dimensionless time, $i$ is the direction of
the noise, and $j$ and $k$ the other two directions. We focus on the
phase flip channel (which means $i=3$) case since completely
equivalent results are found for the other two types of channels.
The corresponding results for bit flip (bit-phase flip) channel are
simply found exchanging $c_{3}$ with $c_{1}$ ($c_{2}$).

As in the Markovian case \cite{Maziero0,Mazzola} one can distinguish
three different dynamical regimes, depending on the relations
between $c_{1}$, $c_{2}$, and $c_{3}$.

\textit{First regime}: $|c_{3}(0)|\geq|c_{1}(0)|,|c_{2}(0)|$.
In this regime, classical correlations remain constant in time. This is
clearly displayed by Eq.~\eqref{cl2}, where $\chi(t)$ is equal
to $|c_{3}(0)|$ which is constant during all the time evolution. The
quantum mutual information characterizing the total correlations
displays damped oscillations, therefore quantum discord is also
exhibiting damped oscillations, tending asymptotically to zero. No
discontinuities appear in quantum or classical correlations
dynamics.

\textit{Second regime}: $c_{3}=0$. Here,  all the three types of correlations
(quantum, classical, total) display damped oscillations and tend
asymptotically to zero, meaning that the state of the system tends
asymptotically to a product state.

\textit{Third regime}: $|c_{3}(0)|<|c_{1}(0)|$ and/or
$|c_{2}(0)|$. This is the region displaying most interesting dynamical features.
Initially,
classical correlations decay till to a certain point in time, and then become
abruptly constant \cite{Maziero}, and at that very same time the
quantum discord changes its decay rate. In Ref.~\cite{Mazzola} we
demonstrated that the initial decay rate can be even equal to zero.
This is mathematically explained observing the analytic formula of
classical correlation in Eq.~\eqref{cl2}: initially $\chi(t)$
is equal to $\max \{|c_{1}(t)|,|c_{2}(t)| \}$ with $c_{1}(t)$ and
$c_{2}(t)$ oscillating functions, then at the sudden change point
$|c_{3}(0)|$ becomes bigger than $\max \{|c_{1}(t)|,|c_{2}(t)| \}$,
therefore $\chi(t)$ is now equal to $|c_{3}(0)|$ which is constant in time, hence classical correlation becomes constant.

As time passes new features may appear in the dynamics of correlations.
In fact if during the time evolution $|c_{1}(t)|$ or $|c_{2}(t)|$
turns bigger than $|c_{3}(0)|$ classical correlation is not constant anymore, and starts again oscillating until it becomes again
abruptly constant. Clearly, sudden changes in the dynamics of
discord take place in correspondence of the discontinuities of
$\chi(t)$. Within this regime we can distinguish two
different behaviours for the dynamics of the quantum correlation: the
frozen behaviour, and the sudden change
dynamics.

\begin{figure}[pb]
\centerline{\psfig{file=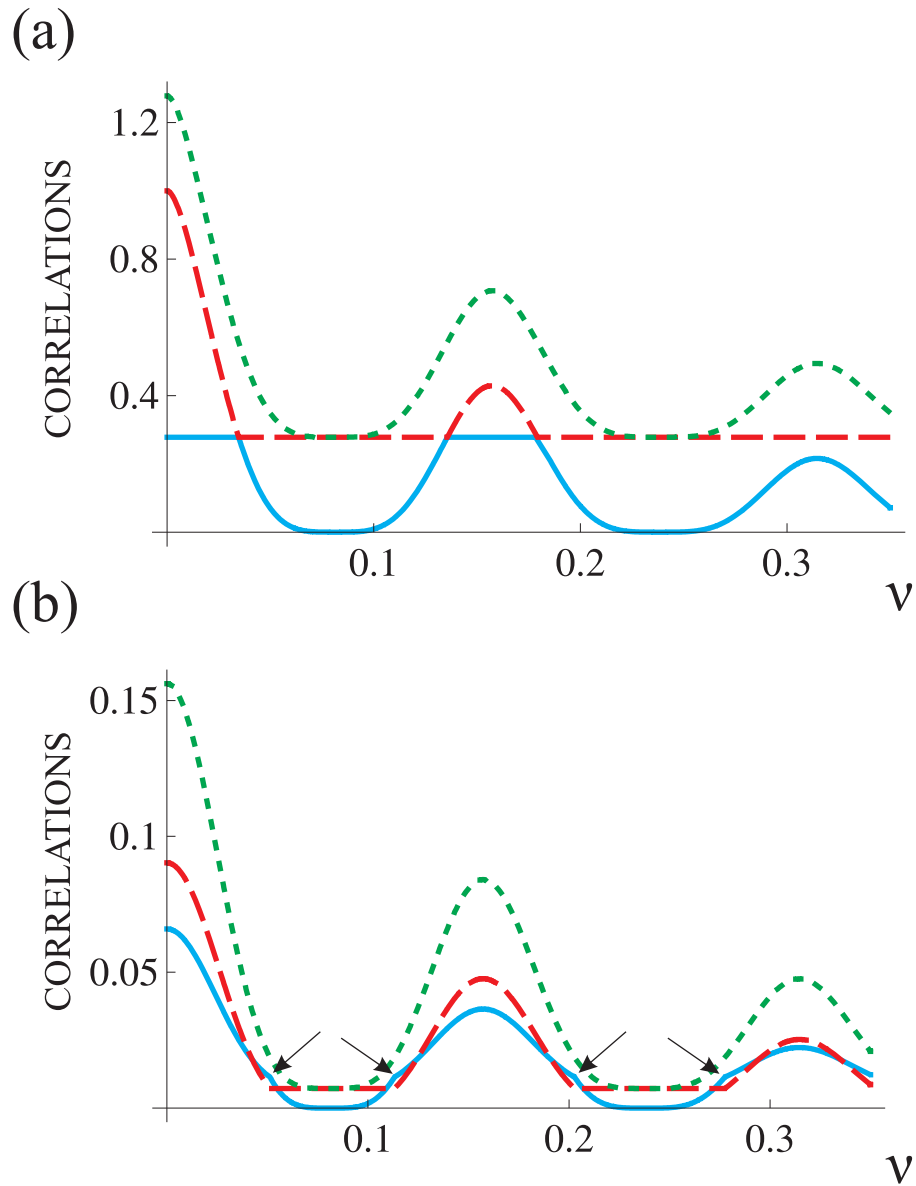}} 
\vspace*{8pt}
\caption{(Color online) Dynamics of mutual information (green
dotted line), classical correlations (red dashed line) and quantum
discord (blue solid line) of two qubits locally interacting with
non-Markovian dephasing channels as a function of scaled time
$\nu=t/2\tau$ with $\tau=5\ s$ and $a=1\ s^{-1}$. (a)
Frozen discord and multiple transition regime,
here we have set $c_1(0)=1$, $c_2(0)=-0.6$ and $c_3=0.6$. (b) Sudden
change regime, obtained setting $c_1(0)=0.35$, $c_2(0)=-0.3$ and
$c_3=0.1$; the arrows emphasize the sudden change points.}\label{figure1}
\end{figure}

(i) Frozen discord with sudden changes displayed in Fig.~\ref{figure1} (a).  The dynamics
displays multiple sudden transitions: classical correlations at first
decrease, while the discord remains constant unaffected by
decoherence, then at a sudden transition point classical correlation
becomes constant and discord oscillates, and then again discord
becomes constant while classical correlation oscillates. The initial state
 in Fig.~\ref{figure1} (a) is of the form
$\rho_{AB}=(1+c_{3})/2\ket{\Psi^{+}}\bra{\Psi^{+}}+(1-c_{3})/2\ket{\Phi^{+}}\bra{\Phi^{+}}$
with $c_{3}=0.6$, however a more general class of states exhibits
frozen discord and (multiple) sudden
transitions. This class of states is of the form of
Eq.~\eqref{maxmixedmarg} where $c_{1(2)}(0)=k$, $c_{2(1)}(0)=-c_{3}k$
and $c_{3}(0)=c_{3}$ with $k$ real and $|k|>|c_{3}|$. For this class
of states the quantum mutual information describing the total
correlation of the system can be cast in a very simple form:
\begin{eqnarray}\label{main}
{\cal I}[\rho_{AB}(t)]&=&\sum_{j=1}^{2}\frac{1+(-1)^{j}
c_3}{2}\log_{2}[1+(-1)^{j}c_3]\\
&+&\sum_{j=1}^{2}\frac{1+(-1)^{j}
c_{1(2)}(t)}{2}\log_{2}[1+(-1)^{j}c_{1(2)}(t)] \nonumber,
\end{eqnarray}
where the first constant term represents either the frozen discord
or the constant classical correlation depending on the dynamical
phase of the system.

(ii) The sudden change regime without frozen discord displayed in Fig.~\ref{figure1} (b).
At first both classical and quantum correlations decrease till, at a
sudden change point (emphasized in the figure by the first arrow on
the left), classical correlation becomes constant while quantum
correlation exhibits a discontinuous change in the amplitude of the damped
oscillations. At the following sudden change point, indicated by the
second arrow, classical correlation starts again to oscillate and
quantum discord changes back to its previous rate of oscillation.

We mention in passing that, for all the regions of parameters,
entanglement exhibits damped oscillations, similar to those of the quantum mutual
information, eventually showing sudden death \cite{yu}.

\section{Geometrical interpretation of frozen discord}
In Ref.~\cite{Modi} Modi \textit{et al.}~propose a new definition
for the correlations of a quantum system. Their idea is to put all
the types of correlations on the same footing defining correlations
in terms of distances (measured by relative entropy) between states.
In analogy to the relative entropy of entanglement, quantum discord
is defined as the distance of the state under study to its closest
classical state, and classical correlation is the distance from
such a classical state to its closest product state.

In general the definition by Modi \textit{et al.}~does not coincide
with the original operational one, however, for Bell-diagonal state
the two formulations lead to the same results. Therefore we can
interpret our results in the light of the correlations as distance
measure approach.

As already noted in Ref. \cite{Mazzola}, given a Bell-diagonal
state
$\rho=\sum_{i=1^4}\lambda_{i}\ket{\Phi_{i}}\bra{\Phi_{i}}$ with
$\ket{\Phi_{i}}$ the Bell states and $\lambda_{i}$, (taken in
decreasing order) the weights of each component, the sudden
transition between classical and quantum decoherence takes place when
the weight $\lambda_{3}$ of component $\Phi_{3}$ becomes larger than
the weight $\lambda_{2}$ of component $\Phi_{2}$. This can be
understood noting that the
expression for the closest classical state to Bell-diagonal states
is
\begin{equation}
\rho_{\rm cl}(t)= \frac{q(t)}{2} \sum_{i=1,2}
\ket{\Psi_i}\bra{\Psi_i}+\frac{1-q(t)}{2}
\sum_{i=3,4}\ket{\Psi_i}\bra{\Psi_i},
\end{equation}
where $q(t)=\lambda_1(t)+\lambda_2(t)$. Therefore when $\lambda_{3}$
becomes bigger than $\lambda_{2}$ the expression of the closest
classical state changes abruptly. It is actually of interest to ask
where those classical states are located in the Hilbert space of the
system, and whether such a class of classical states display a
particular geometry.

\begin{figure}[pb]
\centerline{\psfig{file=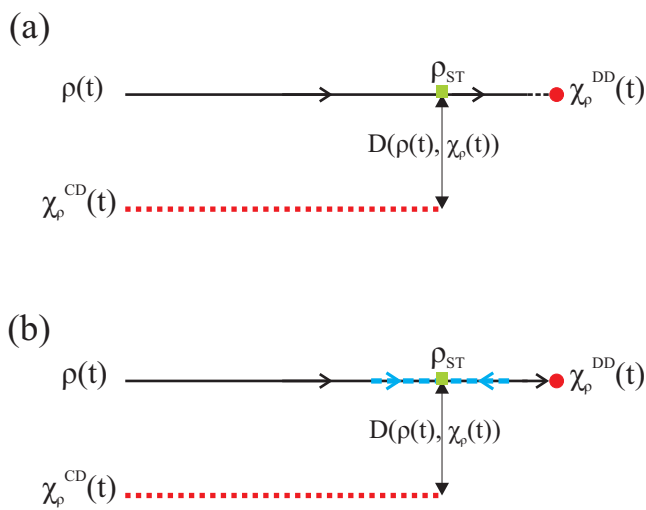}} 
\vspace*{8pt}
\caption{(Color online) Schematization of the trajectories of both the
state of the system and its closest classical state in the Hilbert
space in the (a) Markovian, (b) non-Markovian case.}\label{figure2}
\end{figure}

To fix the ideas we first focus on the Markovian case illustrated in
Fig.~\ref{figure2} (a). There the solid black line represents the
trajectory of the state $\rho(t)$ in a schematized Hilbert space,
and the dotted red line represents the trajectory drawn by the
closest classical state $\chi_{\rho}^{CD}(t)$ (where CD stands for
constant discord). In terms of the distance (relative entropy or
trace distance), this trajectory is parallel to the one traced by
$\rho(t)$ meaning that the discord between the two lines is
constant. The green square on the $\rho(t)$ trajectory indicates the
state $\rho_{ST}$ at which the sudden transition takes place,  in
correspondence of $\lambda_{2}=\lambda_{3}$. The state $\rho_{ST}$
has two different classical states with equal distance: one is the
last state at the end of the red dotted line, i. e.,
$\chi_{\rho}^{CD}(t)$, the other one is indicated by the red sphere
at the right end of the black line $\chi_{\rho}^{DD}(t)$. Since
$\rho(t)$ keeps traveling the black line from left to right, after
the transition $\chi_{\rho}^{DD}(t)$ (DD meaning decaying discord)
becomes the closest classical state. Interestingly
$\chi_{\rho}^{DD}(t)$ is also the asymptotic state of the dynamics
and does not evolve in time.

The non-Markovian case is displayed in Fig.~\ref{figure2} (b),
there the meaning of lines and symbols, and the structure of the set of closest classical states remain the same as in
Fig.~\ref{figure2} (a) but the time evolution of $\rho(t)$ changes.
Due to the non-Markovian memory effects, the state of the system oscillates around the transition point and passes many times from both of the directions
the transition state $\rho_{ST}$. After the first passage of $\rho(t)$
through $\rho_{ST}$ from left to right, $\rho(t)$ hits in a finite time
$\chi_{\rho}^{DD}(t)$, which in the Markovian case was reached only in the
asymptotic limit. Now the memory of the reservoir comes into play
and the state travels back part of the trajectory previously drawn,
this is pictured by the thick dashed blue line partially overlapping
with the black one. This time the state $\rho_{ST}$ is passed from
right to left, so $\rho(t)$ enters again in the constant discord region.
Eventually the direction of the trajectory of $\rho(t)$ in the Hilbert space changes again and the state
enters again the changing-discord-region and remaining there.

\section{Conclusive Remarks}
In this paper we have studied the dynamics of quantum and classical
correlations for a system of two qubits locally interacting with
non-Markovian non-dissipative environments. To solve the dynamics of
the single qubits we extended the model by Daffer \textit{et al.}~to the two-qubit case, when each qubit is subjected to a local colored noise dephasing channel. For initial
Bell-diagonal states the dynamics of correlations exhibits, three different dynamical regimes: (i) constant
classical correlations;  (ii) damped oscillations of quantum and classical correlations (iii) sudden change behavior with possibility for frozen discord including multiple sudden transitions of decoherence. The memory effects of the non-Markovian environment bring
into play non-trivial dynamical features and allow multiple transition points between the classical and quantum decoherence.

We have also proposed a geometrical interpretation of the phenomenon
of frozen discord in terms of trajectory of the
state under investigation and its closest classical state. In the transition point between the classical and quantum decoherence,
the location of the closest classical state changes in discontinuous way. In the Markovian case, this transition point is crossed only once, where as in the non-Markovian case, the memory effects allow several crossings to both directions.

Very recently Jin-Shi et al.~\cite{Jin-ShiNM} have observed the
dynamics of the correlations of two photons in a non-Markovian
dephasing environment and their results display the same features as
described here. In general, we think that results for dephasing
reservoirs shown here and the corresponding experimental results
\cite{Jin-ShiNM} present a considerable step towards more
comprehensive understanding of the dynamics of quantum correlations
in the open system context.

\section*{Acknowledgments}

This work has been supported by the Academy of Finland (Project
No.~133682), the Emil Aaltonen Foundation, the Finnish Cultural Foundation, and the Magnus
Ehrnrooth Foundation. S.M. also thanks
the Turku Collegium of Science and Medicine for financial support.

\end{document}